\documentclass[12pt]{iopart}

\usepackage{graphicx}

\begin{document}

\title[The role of stationarity in magnetic crackling noise]
{The role of stationarity in magnetic crackling noise}

\author{Gianfranco  Durin$^1$ and Stefano Zapperi$^{2,3}$}

\address{$^1$ IEN Galileo Ferraris, strada delle Cacce, 91, 10135 Torino, Italy}

\address{$^2$ CNR-INFM SMC, Dip. di Fisica, Universit\`a
"La Sapienza", P.le A. Moro 2, 00185 Roma, Italy}
\address{$^3$ CNR, Istituto dei Sistemi Complessi, 
Via dei Taurini 19, 00185 Roma, Italy}

\ead{\mailto{durin@ien.it}, \mailto{zapperi@roma1.infn.it}}

\begin{abstract}
We discuss the effect of the stationarity on the avalanche statistics of
Barkhuasen noise signals. We perform experimental measurements on
a Fe$_{85}$B$_{15}$ amorphous ribbon and compare the avalanche distributions
measured around the coercive field, where the signal
is stationary, with those sampled through the entire hysteresis loop.
In the first case, we recover the scaling exponents commonly observed in
other amorphous materials ($\tau=1.3$, $\alpha=1.5$).
while in the second the exponents are  significantly larger
($\tau=1.7$, $\alpha=2.2$). We provide a quantitative explanation
of the experimental results through a model for the depinning
of a ferromagnetic domain wall. The present analysis shed light on
the unusually high values for the Barkhausen noise exponents measured 
by Spasojevic et al. [Phys. Rev. E  54 2531 (1996)].
\end{abstract}

\maketitle

\section{Introduction}

Crackling noise is the typical response of many complex systems
displaying avalanche dynamics \cite{SET-01}. When perturbed by a
slow varying external force or field, these systems respond with
impulsive events which span a broad range of sizes. This noise
occurs in micro- and meso-scopic systems such as superconductors,
ferromagnets, charge density waves, fluids in porous media, fracture,
or plasticity, but also on large-scales, as in the tectonic motion of faults
responsible of earthquakes.

The existence of a such a wide class of phenomena displaying similar
noise features could imply that some general aspects of the dynamics
should always be present, irrespective of the microscopic details
of the system. One possible explanation is that
all these systems are in the proximity of some non-equilibrium
critical point and for this reason they display power law distributions over
several decades, together with $1/f^a$-type noise. To understand
this behavior, statistical mechanics models have been proposed. In some
case, these are able to predict the values of the critical exponents
and provide important tools to analyze and interpret the experimental data
\cite{SET-01}. A significative example is the analysis of the average shape of
the noise pulse \cite{SET-01,KUN-00}, which was predicted theoretically to be
asymmetric, while experiments, for instance in magnetic
systems, show instead a marked leftward asymmetry \cite{DUR-02}.
This discrepancy has been recently resolved, showing that the asymmetry is due to
the negative effective mass of the magnetic objects (i.e. the Bloch domain walls) responsible for the noise \cite{ZAP-05}. This example illustrate how simple
models can improve our understanding of the magnetization dynamics.

A very important feature of crackling noise is the condition of stationarity
during the dynamics. In fact, there
are systems which typically respond with avalanches during a
transient regime, the clearest example being the acoustic emission
prior to fracture in brittle materials \cite{garciamartin97},
but also the magnetic noise emitted during the fracture of steel \cite{KUN-04}.
In other cases, it is not always obvious to understand wether
the system is in a stationary state or not. The question is important because
non-stationarity can bias the scaling of the avalanche distributions.
A notable theoretical example in this respect is offered
by the Random Field Ising Model (RFIM) \cite{SET-05}, where
the critical exponent of the avalanche size distribution $\tau$,
changes in three dimensions from  $\tau=1.6$ to $\tau=2$ in stationary and non-stationary conditions, respectively. The reason behind this result
is that in non-stationary condition the distribution is integrated over
different values of the control parameter, yielding a larger effective
exponent. This mechanism was also discussed in general terms by Sornette in
Ref.~\cite{SOR-94}.

Experimentally the problem of stationarity
has been addressed with a certain detail in the
context of Barkhausen noise, the crackling noise occurring in
ferromagnets as the magnetic field is increased. In 1981, Bertotti and Fiorillo \cite{BER-81}
suggested to investigate the noise properties only in a limited
range of magnetization curve around the coercive field, in order to
ensure the stationarity of the detected noise signal. In this
regime, the magnetization process is ruled by the jerky motion of
domain walls, the interfaces delimiting two domains with
opposite magnetization. Before this fact was pointed out, noise experiments had
been performed all along the hysteresis loop, i.e. in highly non
stationary conditions, leading to results sometimes difficult to interpret
theoretically. This new approach to the investigation of Barkhausen
noise generated a series of important
studies which have highly improved our undestanding of the
phenomenon \cite{ALE-90}. Nevertheless, this scientific  field is still very
active, also because the Barkhausen effect is generally
considered as one of the cleanest examples of crackling noise \cite{SET-01}.
A detailed analysis of the vast scientific literature on the subject can be found
in our recent review \cite{DUR-05}. The main conclusion
is that, despite the large variety of soft magnetic materials,
most of the experimentally measured scaling exponents can
be grouped into two different
universality classes, reflecting the dominating magnetic interactions
\cite{DUR-00}.  Nanocrystalline and polycrystalline materials
are dominated by long range interactions,
originated by stray fields inside the sample. Amorphous materials, 
especially when subjected to an external tensile stress \cite{DUR-99}, are
dominated by the short range elastic tension of the domain wall.

As a matter of fact, there is a number of significative experimental
results which do not fit into these two classes. A few materials show
critical exponents which are within the values found for the two
classes, especially in systems where the domain patters is highly
fragmented, and dominated by residual stresses. This is the often
the case of unstressed amorphous alloys, where the mechanism of
production leaves a high degree of disorder and quenched-in stress
in the material. Thus one may consider these variations as experimental
fluctuations.

A more serious problmem comes from the data reported in the paper of
Spasojevic et al. \cite{SPA-96}, which also employed  an amorphous sample
in an unstressed state, but where the critical exponents are
significantly larger than the ones found in the two classes. This
result has often been referred toas  the signature of another
universality class in soft magnets, although no convincing
theoretical explanation has been proposed. In our opinion, it
is  difficult to believe that this new universality class exists, as it would
require that in the sample employed in  Ref.~\cite{SPA-96}
domain wall dynamics would be ruled by completely different
interactions than in other apparently similar amorphous alloys.
We are more inclined to think that in Ref. ~\cite{SPA-96} the
experimental conditions are significantly different from the
standard ones. Assuming that the noise signal was not stationary,
we can try to explain the observed high values of  critical
exponents. In fact, very similar values of the exponents
were measured earlier in NiFe by sampling the noise over the entire
hystresis loop \cite{LIE-72}.
On the other hand, following the recommendation of Ref.
\cite{BER-81}, the authors of Ref. ~\cite{SPA-96} actually
measured the signal in a small magnetization window around the
coercive field. While this is certainly true, we will show
in this paper that non stationarity can be intrinsic to
the experimental setup and material properties.

To corroborate our hypothesis, we perform a series of
Barkhausen noise measurements on a similar amorphous sample in non stationary
conditions, obtaining critical exponents very close to the ones
reported in Ref. \cite{SPA-96}. Moreover, we also measured the
exponents under stationary conditions, recovering the
values typical of amorphous alloys \cite{DUR-99,DUR-00,URB-95a}.
In addition, we show that there exists at least a single critical exponent which
is not affected by non-stationarity.
This exponent, not determined in Ref. \cite{SPA-96},
relates the average size of an avalanche to
its duration, and was shown in \cite{KUN-00} to coincide with the high frequency exponent of the power spectrum. Hence, we verify that the noise spectrum
does not depend on the stationary state of the dynamics.
Finally, we show that a model of domain wall depinning
\cite{URB-95a,NAR-96,CIZ-97,ZAP-98,BAH-99,QUE-01}
can be used  to explain  quantitatively the scaling exponents for both stationary and non-stationary conditions.

A similar comparison between stationary and non
stationary signals can be very important not only for magnetic systems,
but for any other complex systems showing crackling noise. The lesson
we draw from our analysis is that one need to check carefully for the
stationarity of the signal before a detailed comparison between experiments
and theory can be attempted.

\begin{figure}
\includegraphics{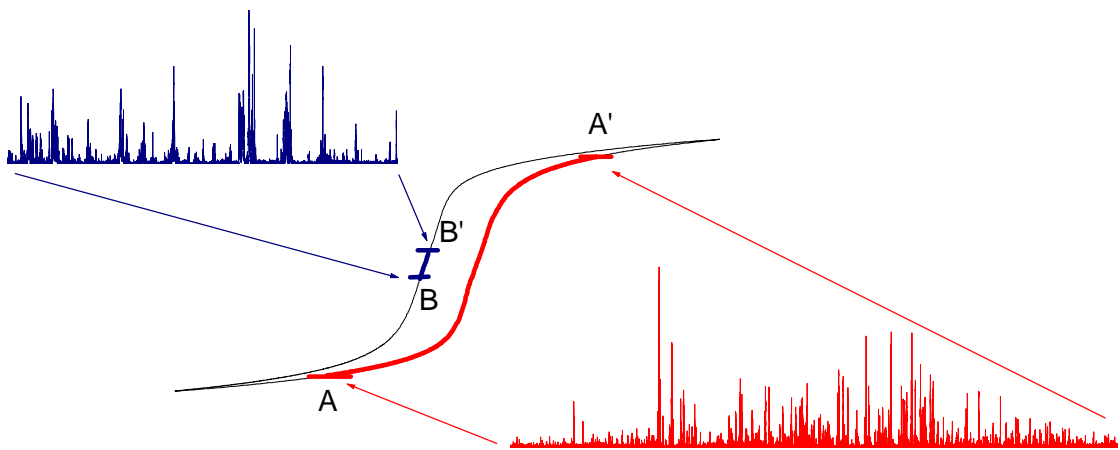}
\caption{Hysteresis loop of an Fe$_{85}$B$_{15}$ amorphous alloy
under moderate tensile stress (10 MPa). The signal detected between
A and A', corresponding to a magnetization range of about 1.8~T, is
highly non stationary, as revealed by the corresponding time signal
in the bottom right corner. The signal between B and B' is
stationary, and corresponds to a range of about 0.1~T.}
\label{fig:loop-and-noise}
\end{figure}

\section{Experimental conditions for stationarity and non-stationarity}

In our cited review \cite{DUR-05}, we have analyzed many
experimental papers of the literature to establish a set of
practical rules to be able to compare experimental data with the theoretical
predictions. A fundamental condition
is to utilize an experimental setup which ensures the stationarity
of the noise signal. This requires to take care of the possible
variations of the experimental conditions during the measurements
and/or inside the materials.

A crucial feature, in this respect, is the role of the demagnetizing
field which is not constant in non-ellipsoidal samples, being is especially
intense close to the ends of the sample. A practical rule is thus
to detect the signal in a region where this field is
reasonably constant, as, for instance, using a pickup coil with
limited width (a few mm, in practice). Another important aspect, as
mentioned in the Introduction, is to delimit the acquired signal to
a small magnetization region around the coercive field. In practice,
this means to consider a region of the hysteresis loop having a
roughly \emph{constant permeability}.
Neither of these two conditions are fulfilled in the paper of
Spasojevic et al. In fact, their amorphous sample is pretty short (4
cm), and thus the the demagnetizing field has a strong spatial
variation which highly affects the magnetization dynamics. In
addition, the \emph{pickup coil is much larger than the sample
length}: this implies that the detected noise is the superposition
of signals of the coils all along the samples, which correspond to
cross sections of the sample having different permeabilities. This
measurement is thus roughly equivalent to consider a material with a
curvy hysteresis loop, and detect the signal all along the loop.

This is exactly the experimental condition we have considered in
order to reproduce results similar to Spasojevic et al obtained for
a Vitrovac amorphous material. In fig.~\ref{fig:loop-and-noise}, we
depict the typical curvy hysteresis loop and the Barkhausen noise in
a highly disordered amorphous ribbon. We have used an
Fe$_{85}$B$_{15}$ high magnetostrictive amorphous alloy, having
dimension of 28 x 1 x 0.002 cm. We have applied a moderate tensile
external stress, as the unstressed sample show a poor
signal-to-noise ratio, and does not permit to determine the critical
exponents with sufficient accuracy. The non stationary signal is
detected  along a quite
extended part of the hysteresis loop, corresponding to about 1.8 T
(A-A' in fig.~\ref{fig:loop-and-noise}), and the applied stress 
is about 10 MPa. This region is visibly much
larger than the linear region of constant permeability around the
coercive field. Thus the stationary signal is detected in a much
smaller magnetization region (B-B' in
fig.~\ref{fig:loop-and-noise}). To improve the signal-to-noise ratio
we have slightly increased the tensile stress to 50 MPa.

The details of the experimental setup have been described at length in
other papers \cite{DUR-00,DUR-05}. Essentially, a 30 cm solenoid
provides a constant applied field all along the sample. We apply a
triangular field with peak amplitude of 70 A/m, reaching a
magnetization of 1 T, smaller than the saturation value. The
coercive field is about 6 A/m. To resolve a well defined sequence of
avalanches, the frequency of applied field is around 10-20 mHz. The
tensile stress is applied using weights of 200 and 1000 g (about 10
and 50 MPa, respectively). 

\section{Experimental results}

The statistical properties of the Barkhausen noise are usually characterized by a 
set distribution functions, typically displaing scaling.
Estimating several scaling exponents in a single material has highly
improved the possibility to verify the theoretical predictions and
compare different systems. In the present experiment we 
will try establish what is the set of 
exponents that depends on the condition of stationarity.

The first step to calculate the distributions is to extract the
sequence of avalanches (or pulses) from the noise signal, introducing a
small threshold which defines the beginning and the end of each
avalanche. The distance between these points is defined as the
avalanche duration $T$, while the size $s$ is the time integral between the
same points. The identification of a small threshold is a procedure which works
pretty well in materials where the signal-to-noise ratio is
relatively large, as for instant in FeCo based amorphous alloys
under tensile stress \cite{ZAP-05,DUR-00}. For more disordered
materials, or with a maze domain structure, as the one used here,
the estimation of the critical exponents is a bit less accurate
because of the reduced signal-to-noise ratio. In any case, we are
interested to see wether there is a significative change of the 
values of the exponents beyond their uncertainty level.

\begin{figure}
\begin{center}
\includegraphics[height=5.6cm]{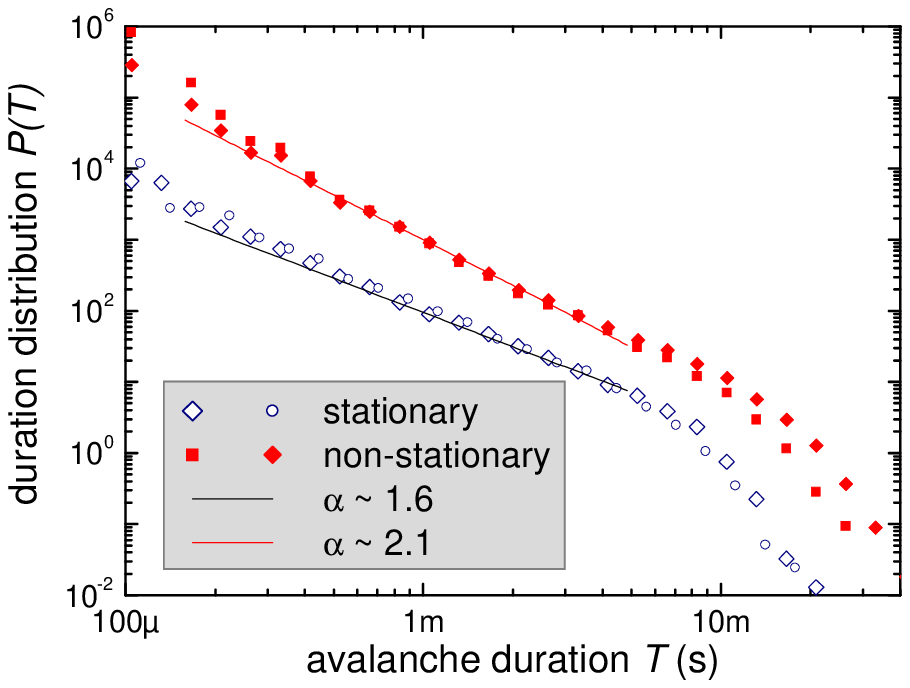}
\hspace{.5cm}
\includegraphics[height=5.5cm]{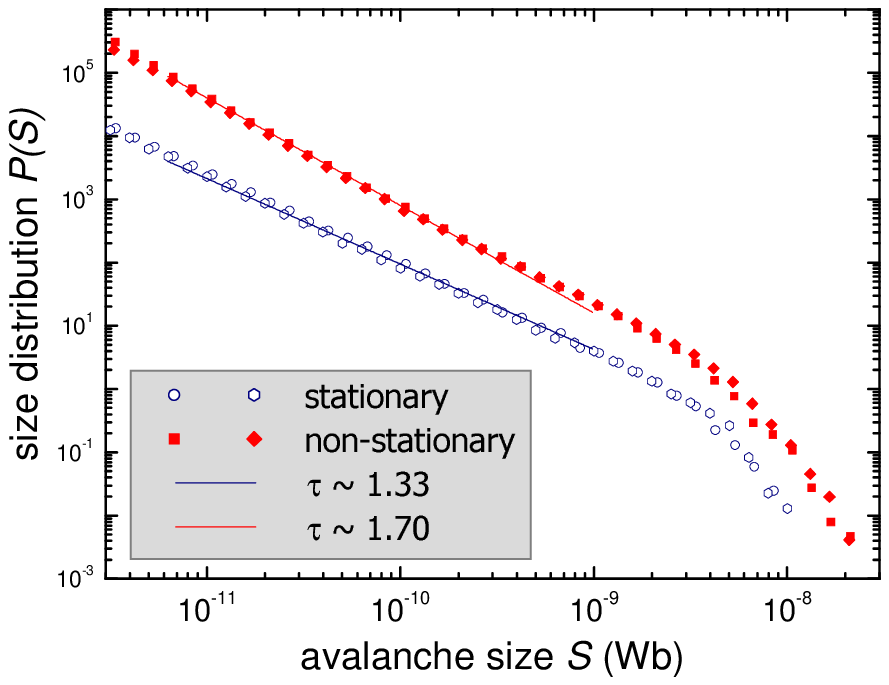}
\caption{Avalanche duration and size distributions for a stationary
(blue) and non-stationary (red) signal. Critical exponents $\alpha$
 and $\tau$ of Eqs.~\ref{eq:pdT-def}-\ref{eq:pds-def} are fitted in the linear part of the plots. Two values of the applied field frequency (10mHz and 20mHz) are
reported, showing that the distributions do not depend on the driving
rate.}
\label{fig:distribs}
\end{center}
\end{figure}

We first consider avalanche duration and size distributions scaling
as
\begin{equation}\label{eq:pdT-def}
P(T)=T^{-\alpha}g(T/T_0)
\end{equation}
and
\begin{equation}\label{eq:pds-def}
    P(S) = S^{-\tau}f(S/S_0)
\end{equation}
where $T_0$ and $S_0$ are the cutoff values, and $\alpha$ and $\tau$ the
critical exponents.
In Fig.~\ref{fig:distribs} we plot these distributions for both
non-stationary and stationary signal, together with an estimate of
the critical exponents obtained by fitting the linear part (in the log-log
scale) of the data. For the non-stationary signal, we get $\alpha =
2.1 \pm 0.1$, and $\tau = 1.70 \pm 0.05$. The larger error for the
duration distribution is justified by the limited range of the data
(about two decades) and the influence of the background noise in the
estimation of shortest avalanches. These values are very close to
the ones reported in ref.~\cite{SPA-96}, as $\alpha = 2.22 \pm
0.08$, and $\tau = 1.77 \pm 0.09$. For completeness, we have also
considered the energy distribution $P(E) \sim E^{-\epsilon}$,
integrating over time of the squared signal, getting $\epsilon =
1.48 \pm 0.05$, comparable with the value od $1.56 \pm 0.05$
obtained in ref.~\cite{SPA-96}.

For the stationary case, we get $\alpha = 1.65 \pm 0.08$, and $\tau
= 1.38 \pm 0.04$. These values are slightly larger than the values
reported for materials belonging to the short range class, where
$\alpha \sim 1.5$, and $\tau \sim 1.30$. It is likely that this
signal is only \emph{nearly} stationary, and a small contribution of
non-stationary avalanches change a little the distribution.

\begin{figure}
\begin{center}
\includegraphics[height=7.85cm]{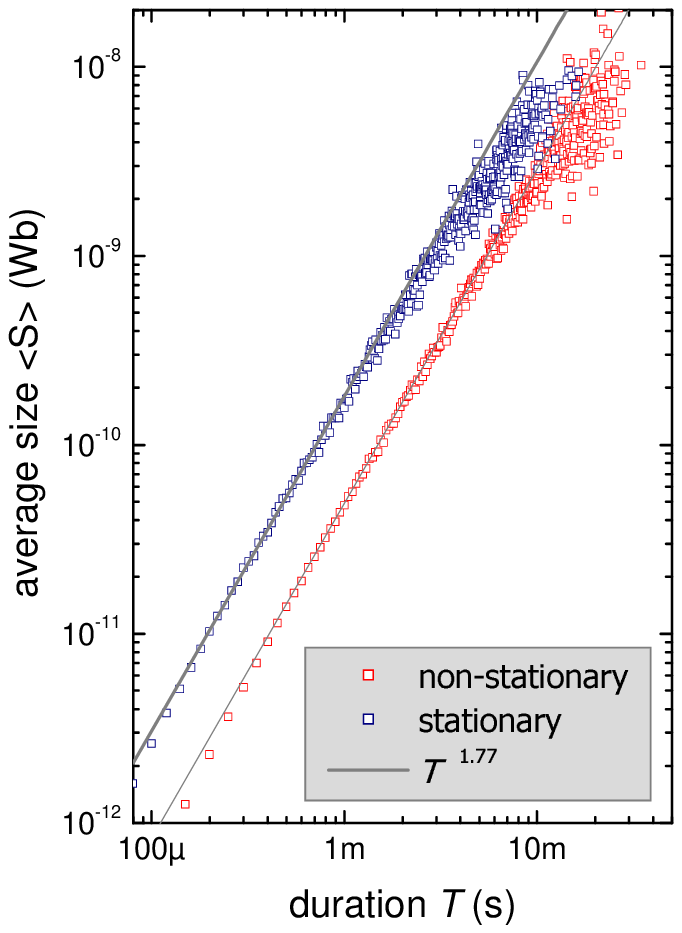}
\hspace{1cm}
\includegraphics[height=8cm]{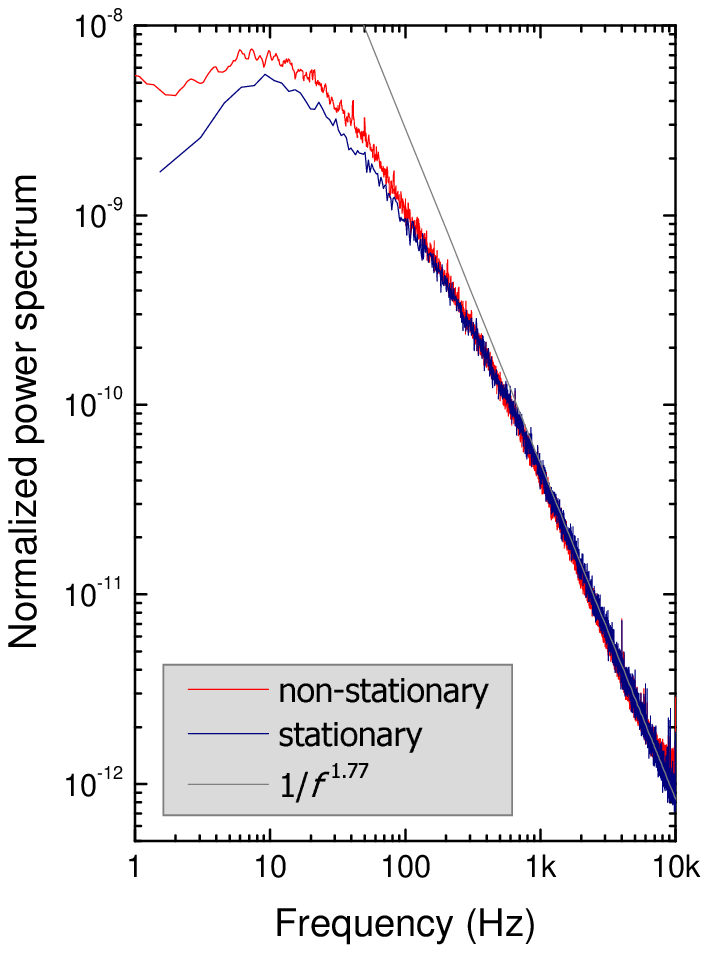}
\caption{(left) Average size of a avalanche as a function of its
duration for stationary and non-stationary signals. From this plot,
it is possible to estimate the exponent $1/\sigma \nu z$ of
eq.~\ref{eq:save_vs_t}. We also plot the theoretical estimation
using the short range universality class ($1/\sigma \nu z \sim
1.77$). (right) The power spectra of the two signals coincide at
high frequency, and are well approximated by $f^{-1/\sigma \nu z}$
}\label{fig:snuzzu}
\end{center}
\end{figure}

We have also considered another important critical exponent, which
relates the \emph{average} avalanche size to its duration
\begin{equation}\label{eq:save_vs_t}
\langle S \rangle \sim T^{1/\sigma \nu z},
\end{equation}
where $\sigma$ is the exponent for the avalanche characteristic
size, $\nu$ is the correlation length  exponent and $z$ is the dynamic
exponent \cite{SET-05}.
The exponent $1/\sigma \nu z$ was shown as well to describe with good
accuracy the high frequency limit of the power spectra
\cite{KUN-00,DUR-02,DUR-05}. In fig.~\ref{fig:snuzzu} we plot both
this distribution and the power spectra for better comparison in
case of stationary and non-stationary signal. Remarkably, the
exponent $1/\sigma \nu z$ is the same and coincides with the
theoretical value of 1.77 expected for amorphous materials \cite{DUR-05}.
At the same time, the two power spectra yields at high
frequency with the same critical exponent. The power spectra have
been normalized by the value of average induced flux rate,
proportional to the average magnetization change in the material.
The differences at low frequencies are simply due to the different
temporal correlations between avalanches, which clearly change in
the case of non-stationary signal.

Unfortunately, Spasojevic et al. did not actually estimated the
exponent $1/\sigma \nu z$, even if they plotted the joint
area-duration distribution. They observed that this distribution is
bounded by two lines with exponents 1.3 and 1.63 
(see Fig. 6 of Ref.~\cite{SPA-96}), but
it is difficult to estimate the exponent $1/\sigma \nu z$ from
this.

\section{Theoretical considerations and numerical simulations}

In order to understand the behavior observed in experiments, we
consider a typical domain wall model \cite{URB-95a,NAR-96,CIZ-97,ZAP-98,BAH-99,QUE-01} and discuss how the
non-stationarity of the noise signal affects the resulting
statistics. We consider a flexible 180$^\circ$ domain wall
separating two regions with opposite magnetization directed along
the $z$ axis. In absence of surface overhangs, we can describe the
position of the domain wall by a single valued function
$h(\vec{r},t)$ . The dynamics of the domain wall is determined
considering the contribution to the local force due to exchange and
magnetocrystalline interactions, magnetostatic and dipolar fields,
and disorder \cite{ZAP-98}. The resulting overdamped equation of
motion is given by \cite{URB-95a,NAR-96,CIZ-97,ZAP-98,BAH-99,QUE-01}
\begin{equation}
\Gamma\frac{\partial h(\vec{r},t)}{\partial t}= H-k\bar{h}+ \gamma_w
\nabla^2h(\vec{r},t) +  +\eta(\vec{r},h),
\label{eq:tot}
\end{equation}
where $\Gamma$ is the eddy current damping constant, $H$ is the
applied field increasing at constant rate, $k$ is an effective
demagnetizing factor, $\bar{h}$ is the center of mass of the wall,
$\gamma_w$ is the domain wall surface tension and $\eta$ is an
uncorrelated Gaussian pinning field. Eq.~(\ref{eq:tot}) has been
shown to quantitatively reproduce the statistical properties of the
Barkhausen noise in amorphous alloys under stress \cite{DUR-99,DUR-00}.

The observed scaling behavior is related to the underlying domain wall
depinning transition, which strictly speaking only occurs for $k=0$.
In this case the wall is pinned unless the external field
overcomes a critical field $H_c$, above which a steady motion ensues.
In general, for interfaces close to the depinning transition, the response to small
variations of the applied field occurs by avalanches whose sizes $S$ are distributed
as
\begin{equation}
P(S)\sim S^{-\tau}f(S (H_c-H)^{1/\sigma},\label{eq:ps}
\end{equation}
indicating that the characteristic avalanche size $S_0$ diverges at the transition
with an exponent $\sigma$.
Similarly the distribution of avalanche durations scales as
\begin{equation}
P(T) \sim T^{-\alpha}g(T(H-H_c)^{1/\Delta}).
\end{equation}
The critical behavior associated to the depinning transition has been
studied using renormalization group methods \cite{NAT-92,NAR-93,LES-97,CHA-01}
and the avalanche exponents can be obtained by scaling relations \cite{ZAP-98}.
In particular the renormalization group
predicts $\tau=1.24$ and $\alpha=1.51$ \cite{DUR-05}
from two-loop $\epsilon=4-d$ expansion
to order O($\epsilon^2$) \cite{CHA-01} while simulations
yield $\tau=1.27$ and $\alpha=1.51$ . The cutoff exponents are approximately given by $1/\sigma=2.2$ and $1/\Delta=1.3$ respectively \cite{DUR-05}. Finally, the power spectrum
exponent is $\gamma=1.77$, equivalent to exponent ruling the scaling
of the size of an avalanche with its duration \cite{DUR-05}.

The discussion above refers to the avalanche statistics sampled at a constant
applied field close to the depinning transition, but this rarely corresponds
to normal experimental conditions for Barkhausen noise measurements
(for a notable exception see the experiments reported in Ref.~\cite{KIM-03}
for thin films). Typically the field is ramped at constant rate
and $k>0$, yielding an effective field $H_{eff}=ct - k\bar{h}$
acting on the domain wall.  In these condition, if
$H_{eff}<H_c$ the domain wall is pinned and the effective field grows until
$H_{eff}>H_c$ when the domain wall start to move increasing the restoring
force provided by the demagnetizing field.  Thus the effective field
remains close to depinning transition and the domain wall displays a stationary avalanche dynamics, with  exponents $\tau$, $\alpha$, $\gamma$ derived above and
cutoffs depending now on $k$ \cite{DUR-00}. These conditions are met for
the stationary signal reported here, as well as for many other experiments
reported in the literature for amorphous alloys \cite{DUR-05}, and indeed
the exponents are in perfect agreement with the theory.

If one would ramp the field slowly and sample the avalanche
distribution over all the values of the {\em effective field}, the
result will be different. This condition would occur if we set $k=0$
and increase the field up to $H_c$, but also for $k>0$ if we
consider the transient regime before the steady state is reached. To
describe this non-stationary regime, we need to integrate
Eq.~\ref{eq:ps} over $H$  obtaining
\begin{equation}
p_{int}(S)=\int^{H_c} dH
S^{-\tau} f(S(H-H_c)^{1/\sigma}) \sim S^{-\tau_{int}},
\end{equation}
with $\tau_{int}=\tau+\sigma$. Using the values of $\tau$ and $\sigma$ reported above, we obtain $\tau_{int}=1.72$. A similar
discussion can be repeated for the avalanche duration distribution, yielding
$\alpha_{int}=\alpha+\Delta=2.25$. The exponent $\gamma$ will of course not
resent by the integration and hence should be the same both for stationary
and non-stationary conditions.

\begin{figure}
\includegraphics[width=8cm]{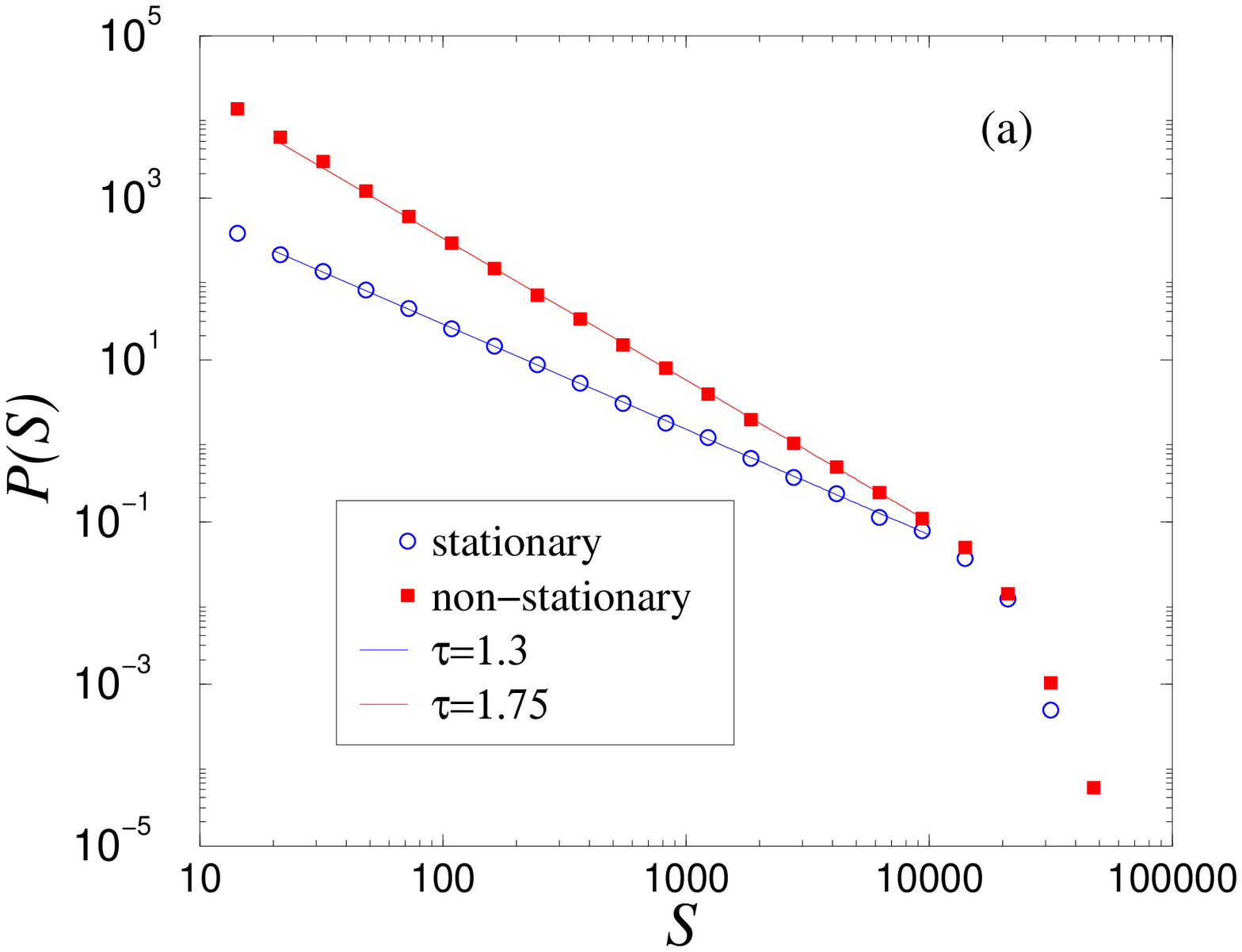}\includegraphics[width=8cm]{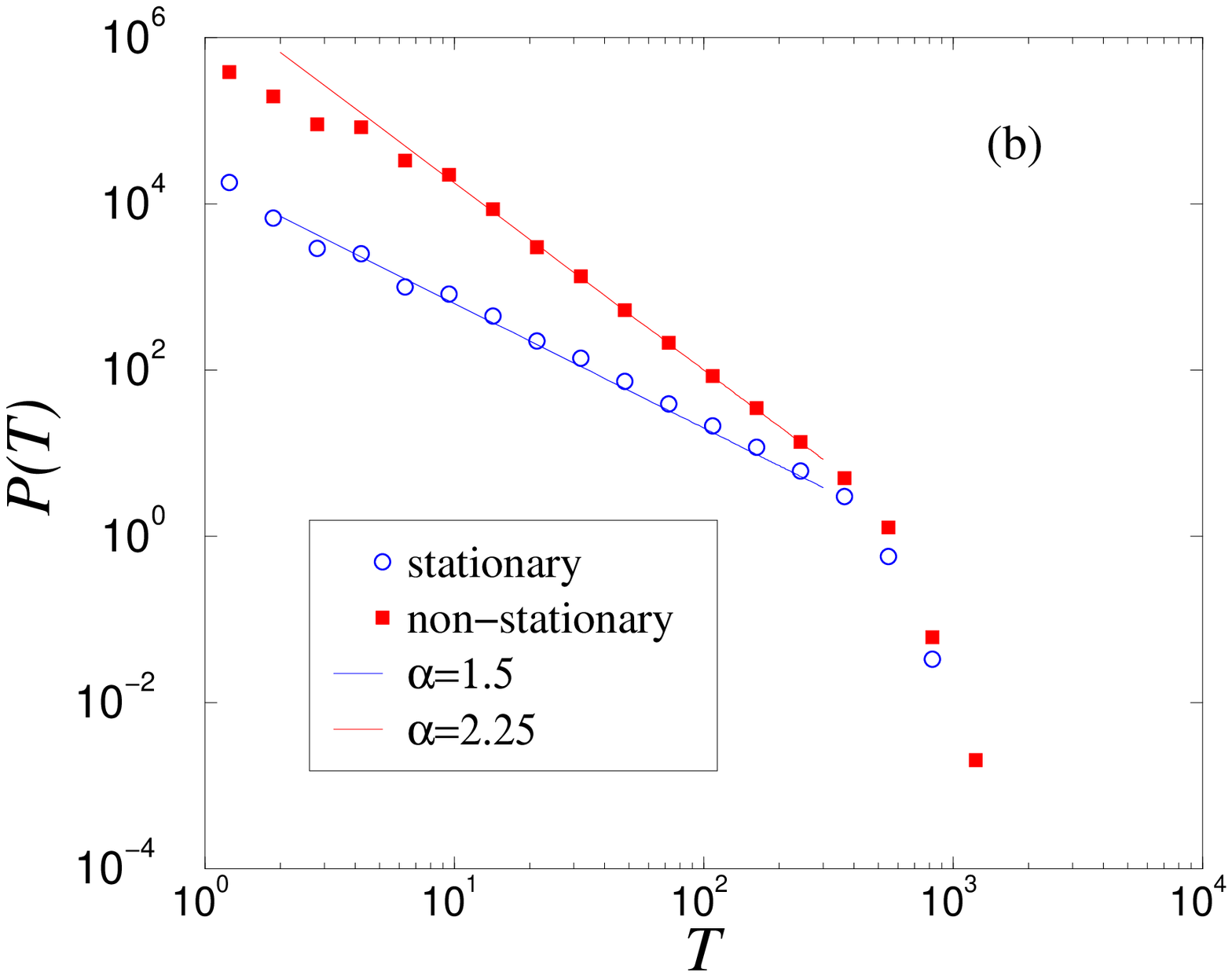}
\caption{The avalanche distributions from numerical simulations performed in stationary and non-stationary conditions: (a) Size distribution. (b) Duration distribution.}\label{fig:pspt_simul}
\end{figure}

\begin{figure}
\includegraphics[width=8cm]{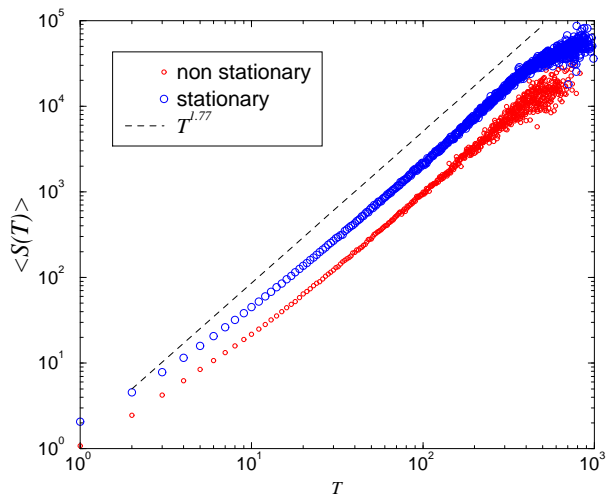}
\caption{Average size of a avalanche as a function of its
duration for stationary and non-stationary signals obtained from numerical simulations.}\label{fig:st_sim}
\end{figure}

From the discussion above we can conclude that the differences
between stationary and non-stationary avalanche signals is simply
due to the integration of the scaling function in the latter case.
As a further illustration, we have simulated an automaton version of
Eq.~(\ref{eq:tot}) and computed the avalanche distributions in the
steady-state regime and in the initial transient. As shown in
Fig.~\ref{fig:pspt_simul}, the scaling exponents reproduce
quantitatively the experimental results. In particular, we find
$\tau=1.3$ and $\alpha=1.5$ in stationary conditions and $\tau=1.75$
and $\alpha=2.25$ for non-stationary conditions. Fig.~\ref{fig:st_sim}
shows as welll that $1/\sigma\nu z$ is does not depend on the
stationarity if the signal.

\section{Conclusions}

We have measured the avalanches distributions in an amorphous
material with high quenched-in disorder, performing a comparitive analysis 
of stationary and non-stationary signals. We have shown that
the size and duration distributions are highly affected by
non-stationarity, while the scaling relation between the average
size and duration is a robust quantity. This result is
actually non surprising, since the non-stationarity only changes the
probability of having an avalanche (and thus the overall
statistics), but does not affect its internal structure. 

The present discussion can be extended to other
classes of crackling noise where problems of non-stationarity
may arise. From our analysis the exponent $1/\sigma\nu z$ emerges 
as the most reliable test for the universality class 
of crackling noise statistics, while one should be aware of the
possible bias in the distribution exponents.
For the Barkhausen noise case, we have provided a quantitative 
explanation of the experiments
by analyzing a model for domain wall depinning. The exponents 
predicted theoretically for the model in stationary and non-stationary
conditions reproduce to a very good precision the experimental 
measurements. A similar resoning would apply to other kind of critical
phenomena as discussed in Ref.~\cite{SOR-94}.

Finally, we think to have clearly demonstrated that the experiment
of Spasojevic et al. \cite{SPA-96} does not imply the existence of a third
universality class in the Barkhausen noise, but only that the results
were probably obtaining under non-stationary conditions. A similar 
discussion applies to the earlier measurements reported in Ref.~\cite{LIE-72},
where the distributions were sampled along the entire hysteresis loop.

\section*{Acknowledgments}
SZ thanks for hospitatility the Kavli Institute for Theoretical Physics, UCSB,
where this work was completed and acknowledge partial financial support through
NSF Grant PHY99-07949.

\section*{References}
\providecommand{\newblock}{}

\bibliographystyle{iopart-num}


\end{document}